\documentclass[traditabstract]{aa}

\usepackage{amsmath}
\usepackage{amsfonts}
\usepackage{txfonts}
\usepackage{amssymb}
\usepackage{graphicx}
\usepackage{natbib}
\begin{document}

\newcommand{\abs}[1]{\ensuremath{\left\lvert #1\right\rvert}}
\newcommand{\erf}[2][]{\ensuremath{\text{erf#1}\left(#2\right)}}
\newcommand{\kB}{\ensuremath{k\!_B}}
\newcommand{\eps}{\ensuremath{\varepsilon}}
\newcommand{\lD}{\ensuremath{\lambda_{\text D}}}

\newcommand{\be}{\begin{equation}}
\newcommand{\ee}{\end{equation}}
\newcommand{\bs}{\begin{subequations}}
\newcommand{\es}{\end{subequations}}
\def\ee{\end{equation}}
\def\be{\begin{equation}}
\def\bdm{\begin{displaymath}}
\def\edm{\end{displaymath}}
\def\noi{\noindent}
\def\ep {\epsilon }
\def\es {\epsilon _s}
\def\noi{\noindent }
\def\l{\left}
\def\r{\right}
\def\a{\alpha }
\def\kr{cosmic ray }
\def\om{\omega }
\def\omr{\omega _R}
\def\Om{\Omega _a}
\def\wa{\omega _{p,a}}
\def\wpe{\omega _{p,e}}
\def\upa{u_{\parallel,a}}
\def\2ku{\sqrt{2}ku_{\parallel, a}}
\def\a2ku{\sqrt{2}|k|u_{\parallel, a}}
\def\Zst{Z^{'}}


\newcommand{\R}{\ensuremath{\mathds{R}}}

\newcommand{\pa}{\ensuremath{_\shortparallel}}
\newcommand{\se}{\ensuremath{_\perp}}

\newcommand{\Th}{\ensuremath{\varTheta}}
\newcommand{\Ga}{\ensuremath{\varGamma}}
\newcommand{\La}{\ensuremath{\varLambda}}

\newcommand{\uint}{\ensuremath{\int_{-\infty}^\infty}}

\newcommand{\f}[1]{\ensuremath{\boldsymbol{#1}}}
\newcommand{\im}{\ensuremath{\mathfrak{I\!m}}}

\newcommand{\pd}[2][]{\ensuremath{\frac{\partial #1}{\partial #2}}}
\newcommand{\df}{\ensuremath{\mathrm{d}}}

\newcommand{\etal}{ \emph{et\,al.}}

\title{Klein-Nishina steps in the energy spectrum \\of galactic cosmic ray electrons}
\author{R. Schlickeiser\thanks{\emph{e-mail:} rsch@tp4.ruhr-uni-bochum.de} 
  \and J. Ruppel\thanks{\emph{e-mail:} jr@tp4.ruhr-uni-bochum.de}}
\institute{Institut f\"ur Theoretische Physik, Lehrstuhl IV:
Weltraum- und Astrophysik, Ruhr-Universit\"at Bochum,
D-44780 Bochum, Germany}
\date{\today}

\abstract{ The full Klein-Nishina cross section for the inverse
  Compton scattering interactions of electrons implies a significant
  reduction of the electron energy loss rate compared to the Thomson
  limit when the electron energy exceeds the critical Klein-Nishina
  energy $E_K=\gamma _Km_ec^2=0.27m^2_ec^2/(k_BT)$, where $T$ denotes
  the temperature of the photon graybody distribution. As a
  consequence the total radiative energy loss rate of single electrons
  exhibits sudden drops in the overall $\dot{\gamma }\propto \gamma
  ^2$-dependence when the electron energy reaches the critical
  Klein-Nishina energy. The strength of the drop is proportional to
  the energy density of the photon radiation field.  The diffuse
  galactic optical photon fields from stars of spectral type B and G-K
  lead to critical Klein-Nishina energies of 40 and 161 GeV,
  respectively. Associated with the drop in the loss rate are sudden
  increases (Klein-Nishina steps) in the equilibrium spectrum of \kr
  electrons. Because the radiative loss rate of electrons is the main
  ingredient in any transport model of high-energy \kr electrons,
  Klein-Nishina steps will modify any calculated electron equilibrium
  spectrum irrespective of the electron sources and spatial transport
  mode. To delineate most clearly the consequences of the
  Klein-Nishina drops in the radiative loss rate, we chose as
  illustrative example the simplest realistic model for \kr electron
  dynamics in the Galaxy, consisting of the competition of radiative
  losses and secondary production by inelastic hadron-hadron
  collisions. We demonstrate that the spectral structure in the FERMI
  and H.E.S.S. data is well described and even the excess measured by
  ATIC might be explained by Klein-Nishina steps.}


\maketitle

\titlerunning{Klein-Nishina steps in the energy spectrum of galactic
  cosmic ray electrons} \authorrunning{R. Schlickeiser \& J. Ruppel}

\section{Introduction}
Recent measurements of the energy spectrum of local galactic \kr
electrons at energies above a few hundred GeV by the ATIC instrument
\cite{at08} have reported a significant excess in the all-electron
intensity that agrees at lower energies with the measurements of the
PAMELA satellite experiment \cite{pa09}, which also has observed a
dramatic rise in the positron fraction starting at 10 GeV and
extending up to 300 GeV. The significant ATIC excess has not been
confirmed by the electron spectrum determinations with the FERMI
satellite \cite{fe09} and the H.E.S.S. air Cherenkov \cite{ah08}
experiments, although these measurements indicate some spectral
structure deviating from a pure power law behaviour in the ATIC energy
range.  These observations have motivated a large number of
interpretations, from possible signatures of dark matter annihilation
(e.g. \cite{sh09}) to nearby astrophysical electron sources.

Here we explain the ATIC excess by a classical effect which sofar has
not been discussed in this context: during their galactic propagation
positrons and electrons with energies above 10 GeV are subject to
synchrotron radiation losses in the galactic magnetic field of about
$3\mu $G and inverse Compton radiation losses in galactic target
photon fields listed in Table \ref{table_rf}, including the universal
microwave background radiation field, infrared photons and optical
stellar photons. The diffuse galactic optical photons can be
characterized by the superposition of two graybody distributions (see
the discussion in section 2.3 of \cite{s02}):

\noi 1) photons from stars of spectral type G-K with energy density
$W_G=0.3$ eV cm$^{-3}$ and temperature $T_G=5000$ K, corresponding to
a mean photon energy $<\epsilon >_G=2.7k_BT_G=2.327\cdot
10^{-4}T_G=1.16$ eV;

\noi 2) photons from stars of spectral type B with energy density
$W_B=0.09$ eV cm$^{-3}$ and temperature $T_B=20000$ K, corresponding
to a mean photon energy $<\epsilon >_B=2.7k_BT_G=2.327\cdot
10^{-4}T_B=4.65$ eV.

For electron Lorentz factors $\gamma $ much smaller than the critical
Klein-Nishina Lorentz factor $\gamma _{K}=0.27m_ec^2/k_BT=1.58\cdot
10^9/T(K)$ (see Eq. (\ref{a3}) below), the inverse Compton scattering
cross section of a single electron can be well approximated by the
Thomson cross section resulting in the standard energy loss rate of
single electrons $\dot{\gamma }=-4c\sigma _TW\gamma ^2/(3m_ec^2)$,
where $\sigma _T=6.65\cdot 10^{-25}$ cm$^2$ denotes the Thomson cross
section and $c$ the speed of light. However, for Lorentz factors
$\gamma \ge \gamma _{KN}$ the full Klein-Nishina cross section has to
be used \cite{j65,bk70,s79,ps79} resulting in a significant reduction
of the inverse Compton loss rate. For the two graybody optical photon
distributions the respective critical Klein-Nishina Lorentz factors
are $\gamma _{KN,G}=3.2\cdot 10^5$, corresponding to an electron
energy of $E_{KN,G}=161$ GeV, and $\gamma _{KN,B}=7.9\cdot 10^4$,
corresponding to an electron energy of $E_{KN,G}=40$ GeV.  We will
demonstrate that this Klein-Nishina reduction of the inverse Compton
energy loss rate leads to Klein-Nishina steps in the \kr electron
equilibrium spectrum which describes the observed FERMI and
H.E.S.S. data well.  In Sect. 2 we determine the galactic synchrotron
and inverse Compton energy loss rates in the full Klein-Nishina
case. For the illustrative example of a purely secondary origin of
galactic electrons we show in Sect. 3 the resulting Klein-Nishina
steps in comparison with the recent electron spectrum observations.
\begin{table}
  \label{table_rf}
  \caption{Electromagnetic graybody radiation fields in the local interstellar medium}
  \centering
  \begin{tabular}{l l l l l l}
    \hline
    i & comment & $T_i$ / K & $W_i$ / $\frac{\textnormal{eV}}{\textnormal{cm}^3}$ & $\gamma_{K,i}$ & $E_{K,i}$ / GeV\\
    \hline \hline
    1 & spectral type B & 20000 & 0.09 & $7.9 \cdot 10^4$ & 40\\
    2 & spectral type G - K &   5000 & 0.3   & 3.2$\cdot 10^5$ & 161\\
    3 & infrared &  20 & 0.4    & 7.9$\cdot 10^7$ & 4.0$\cdot 10^4$\\
    4 & microwave &  2.7 & 0.25 & 5.9$\cdot 10^8$ & 3.0$\cdot 10^5$\\
    \hline
  \end{tabular}
\end{table}

\section{Synchrotron and inverse Compton energy loss rates}
The synchrotron energy loss rate of a single electron in a large-scale
random magnetic field of constant strength $B$ is \cite{cs88}
\be
|\dot{\gamma }|_S={4\sigma _Tc\over 3m_ec^2}U_B\gamma ^2~,
\label{a1}
\ee
where $U_B=B^2/8\pi =0.22b^2_3$ eV cm$^{-3}$ if we scale the galactic
magnetic field strength as $B=3b_3´\mu $G.

In the Appendix we approximately calculate the inverse Compton energy
loss rate of a single electron in one graybody photon field as
\be
|\dot{\gamma }|_{\rm C}\simeq {4\sigma _TcW\over 3m_ec^2}{\gamma _{K}^2\gamma ^2\over \gamma _K^2+\gamma ^2}~,
\label{a2}
\ee
where the critical Klein-Nishina Lorentz factor is given by 
\be
\gamma _{K}\equiv {3\sqrt{5}\over 8\pi }{m_ec^2\over k_BT}={0.53m_ec^2\over k_BT}~.
\label{a3}
\ee
For small electron Lorentz factors $\gamma \ll \gamma _{K}$ the
general inverse Compton energy loss rate (\ref{a2}) reduces to the
Thomson limit
\be
|\dot{\gamma }|_{\rm C}\l(\gamma \ll \gamma _{K}\r)\simeq {4\sigma _TcW\over 3\pi ^4m_ec^2}\gamma ^2~,
\label{a4}
\ee
whereas for large electron Lorentz factors $\gamma \gg \gamma _{K}$ we
obtain the energy-independent extreme Klein Nishina limit
\be
|\dot{\gamma }|_{\rm C}\l(\gamma \gg \gamma _{KN}\r)\simeq {4\sigma _TcW\over 3m_ec^2}\gamma _{K}^2~.
\label{a5}
\ee
The total radiative (synchrotron and inverse Compton) energy loss rate
of a single electron is given by the sum of rate (\ref{a1}) and rates
(\ref{a2}) for the four diffuse galactic radiation fields listed in
Table \ref{table_rf}, yielding
\be
|\dot{\gamma }|_{\rm R}={4\sigma _TcU_B\gamma ^2\over 3m_ec^2}
\l[1+\sum _{i=1}^4{W_i\over U_B}{\gamma ^2_{K,i}\over \gamma ^2+\gamma ^2_{K,i}}\r]~.
\label{a6}
\ee
\begin{figure}
  \setlength{\unitlength}{0.00044\textwidth}
  \begin{center}
    \begin{picture}(1100,1090)(-80,-50)
      \put(-70,440){\rotatebox{90}{\large$\log_{10} \left(|\dot{\gamma }|_{\rm R} / \gamma^{2}\right)$}}%
      \put(400,-70){\large$\log_{10} (E_e / \textnormal{GeV})$}%
      \includegraphics[width=1000\unitlength]{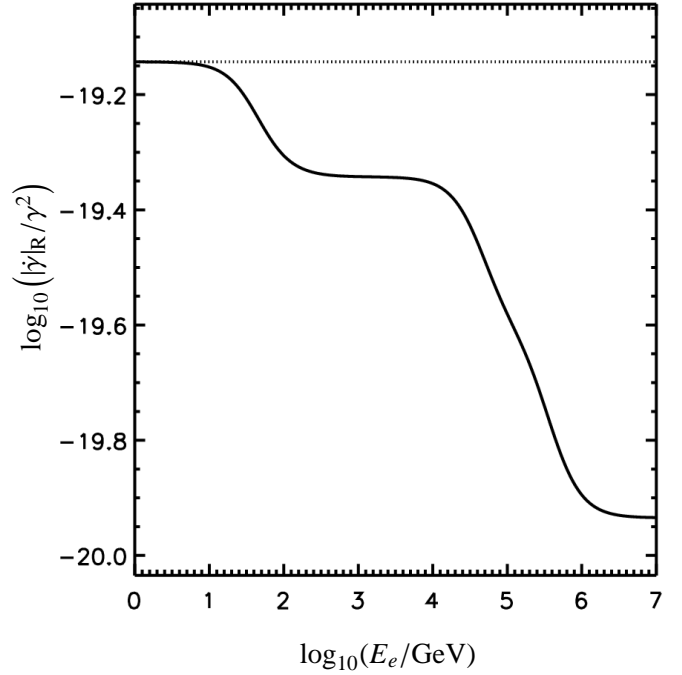}
    \end{picture}
  \end{center}
  \caption{
  \label{energy_losses_fig}
  The energy loss of relativistic electrons -- as given by equation
  (\ref{a6}) -- as a funtion of the electron energy in GeV. The solid
  line shows four drops related to the critical Klein-Nishina
  energies, whereas the dotted line illustrates synchrotron and
  inverse Compton losses in the Thomson limit.}
\end{figure}
In Figure \ref{energy_losses_fig} we show the resulting radiative
energy loss rate for the local galactic magnetic field and photon
energy densities for relativistic electrons with energies between 1
and $10^7$ GeV. One clearly notices the four sudden drops whenever the
electron energy reaches each of the critical Klein-Nishina
energies. The strength of the drop is proportional to the energy
density of the photon field. For electron Lorentz factor below the
smallest critical Klein-Nishina Lorentz factor all four graybody
photon fields plus the magnetic field energy density contribute to the
loss rate. Once the electron Lorentz factor has exceeded the critical
Klein-Nishina Lorentz factor of a particular photon field, this photon
field no longer contributes to the radiative loss rate due to the much
reduced inverse Compton loss rate in the Klein-Nishina limit. At
Lorentz factors above the maximum critical Lorentz factor from the
microwave background photons $\gamma _{k,4}=5.9\cdot 10^8$ only
synchrotron losses contribute to the radiative loss rate.
\section{Klein-Nishina steps in the electron equilibrium spectrum}
In this section we calculate the equilibrium spectrum of galactic \kr
electrons above 10 GeV taking into account the modified radiative loss
rate (\ref{a6}), as well as non-thermal bremsstrahlung, adiabative
deceleration losses in a possible galactic wind with the velocity
$v_{gw}$ and Coulomb and ionization losses\cite{p93}. Because the
radiative loss rate of electrons is the main ingredient in any
transport model of high-energy \kr electrons, Klein-Nishina steps will
modify any calculated electron equilibrium spectrum irrespective of
the electron sources and spatial transport mode. To delineate most
clearly the consequences of the Klein-Nishina drops in the radiative
loss rate, we chose as illustrative example the simplest realistic
model for \kr electron dynamics in the Galaxy. Extensions to more
sophisticated models of \kr electron dynamics (influence of localized
point sources, spatial diffusion, convection and distributed
reacceleration), where the consequences of the modified inverse
Compton losses also occur, are the subject of future work.

At electron energies above 10 GeV the electron's radiative loss time
$\tau _R=\gamma /|\dot{\gamma }_R|\propto \gamma ^{-1}$ is so short
that the Galaxy behaves as a thick target or fractional
calorimeter\cite{v89,p93} for the electrons. The equilibrium energy
spectrum of \kr electrons $N(\gamma )$ then results from the balance
of electron production, expressed as injection spectrum $Q(\gamma )$,
and radiative energy losses from the solution of the balance equation
\be
{d\over d\gamma }\l[|\dot{\gamma }_{\rm R}(\gamma )|N(\gamma )\r]+Q(\gamma )=0,
\label{c1}
\ee
implying
\be
N(\gamma )=|\dot{\gamma }_{\rm R}(\gamma )|^{-1}\int_{\gamma }^\infty dy\, Q(y)~.
\label{c2}
\ee

Moreover, we assume here that all electrons are secondaries resulting
from inealastic hadron-hadron collisions of primary \kr hadrons with
interstellar gas atoms and molecules during their confinement in the
Galaxy. It is well established\cite{lio05,del09} that secondary
production accounts for the major part of the observed galactic \kr
electrons at relativistic energies. The locally measured hadron
spectrum\cite{an04} at energies below $4.4\cdot 10^{15}$ GeV is a
power law $\propto \gamma_h^{-s}$, with spectral index $s=2.74$.
Using the hadron-hadron cross section templates\cite{ka06} the
resulting electron injection spectrum at energies above 10 GeV,
$Q(\gamma )=Q_0\gamma ^{-s}$, then follows a power law with the hadron
spectral index $s$. With this injection spectrum and the radiative
energy loss rate (\ref{a6}) the equilibrium spectrum (\ref{c2})
becomes
\begin{align}
  N(\gamma )&={Q_0\gamma ^{1-s}\over (s-1)|\dot{\gamma }|_{\rm R}(\gamma )|}
  \nonumber\\
  &={3m_ec^2Q_0\gamma ^{-s-1}\over 4(s-1)\sigma _TcU_B}
  \l[1+\sum _{i=1}^4{W_i\over U_B}{\gamma ^2_{K,i}\over \gamma ^2+\gamma ^2_{K,i}}\r]^{-1}
  \label{c3}
\end{align}
which is shown in Figure \ref{data} in comparison with the observed
energy spectrum of galactic \kr electrons. It can be seen that the
spectral shape of the H.E.S.S. and FERMI data is well fitted. In
Figure \ref{comparison} the Klein-Nishina steps are clearly visible.
\begin{figure}[ht]
  \setlength{\unitlength}{0.00044\textwidth}
  \begin{center}
    \begin{picture}(1100,1079)(-80,-50)
      \put(-70,170){\rotatebox{90}{\large$\log_{10} (E^{k}\,N_e(E_e) / $GeV$^{k-1}$m$^{-2}$s$^{-1}$sr$^{-1})$}}%
      \put(400,-70){\large$\log_{10} (E_e / \textnormal{GeV})$}%
      \includegraphics[width=1000\unitlength]{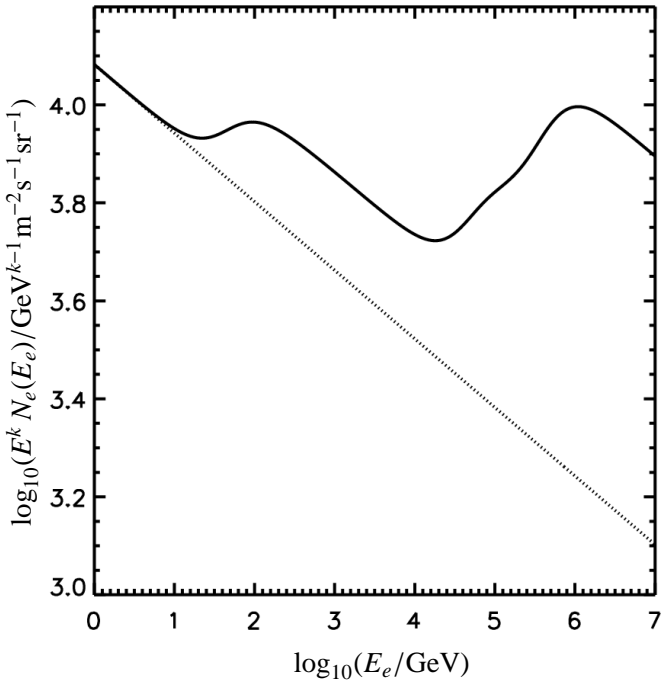}
  \end{picture}
  \end{center}
  \caption{
  \label{comparison}
  The equilibrium spectrum of cosmic ray electrons as a function of
  the electron energy in GeV. The solid line represents the electron
  spectrum with Klein-Nishina corrections, the dotted line shows the
  spectrum in the Thomson limit. The comparison of the shapes clearly
  illustrates the impact of the Klein-Nishina corrections.}
\end{figure}
\begin{figure}[ht]
  \setlength{\unitlength}{0.00044\textwidth}
  \begin{center}
    \begin{picture}(1100,1079)(-80,-50)
      \includegraphics[width=1000\unitlength]{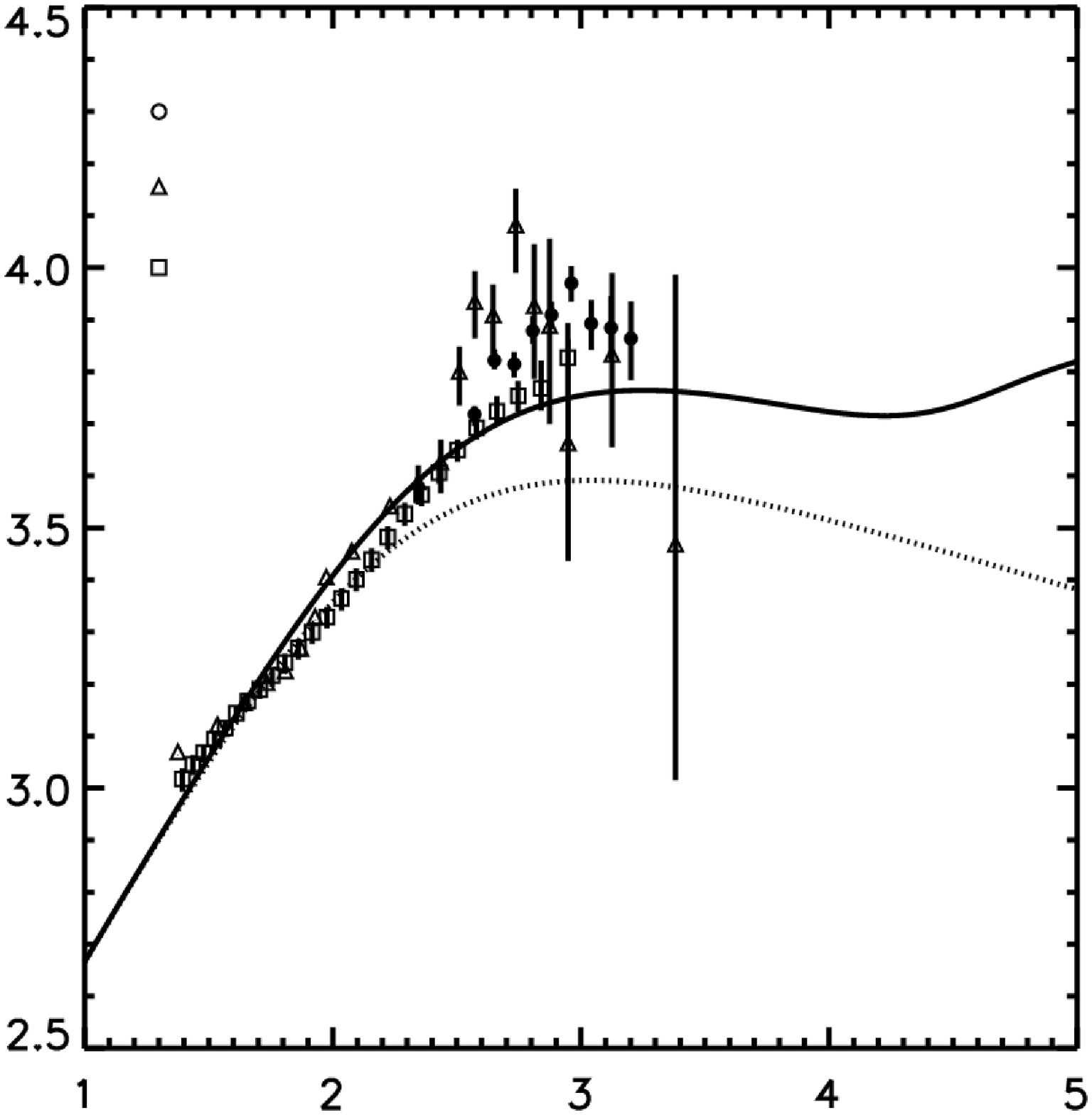}
      \put(-1070,170){\rotatebox{90}{\large$\log_{10} (E^{k}\,N_e(E_e) / $GeV$^{k-1}$m$^{-2}$s$^{-1}$sr$^{-1})$}}%
      \put(-600,-70){\large$\log_{10} (E_e / \textnormal{GeV})$}%
      \put(-825,910){H.E.S.S.}%
      \put(-825,840){ATIC}%
      \put(-825,765){FERMI}%
  \end{picture}
  \end{center}
  \caption{
  \label{data}
  The equilibrium spectrum of cosmic ray electrons as a function of
  the electron energy in GeV compared to observational data by
  H.E.S.S., ATIC and FERMI. Again, the solid line represents the
  electron spectrum with Klein-Nishina corrections, the dotted line
  shows the spectrum in the Thomson limit.  Whereass here, the full
  energy losses are taken into account. The values of the parameters
  are: \mbox{$B=3 \, \mu $G}, \mbox{$n_H=0.3$\,cm$^{-3}$}, Degree of
  ionisation: 0.2, \mbox{$k=3.6$},
  \mbox{div$(v_{gw})=10^{-13.13}$\,s$^{-1}$}}
\end{figure}
\section {Summary and conclusions}
The full Klein-Nishina cross section for the inverse Compton
scattering interactions of electrons implies a significant reduction
of the electron energy loss rate compared to the Thomson limit when
the electron energy exceeds the critical Klein-Nishina energy
$E_K=\gamma _Km_ec^2=0.27m^2_ec^2/(k_BT)$, where $T$ denotes the
temperature of the photon graybody distribution. As a consequence the
total radiative energy loss rate of single electrons exhibits sudden
drops in the overall $\dot{\gamma }\propto \gamma ^2$-dependence when
the electron energy reaches the critical Klein-Nishina energy. The
strength of the drop is proportional to the energy density of the
photon radiation field.  The diffuse galactic optical photon fields
from stars of spectral type B and G-K lead to critical Klein-Nishina
energies of 40 and 161 GeV, respectively. Associated with the drop in
the loss rate are sudden increases (Klein-Nishina steps) in the
equilibrium spectrum of \kr electrons (see Figure
\ref{comparison}). Because the radiative loss rate of electrons is the
main ingredient in any transport model of high-energy \kr electrons,
Klein-Nishina steps will modify any calculated electron equilibrium
spectrum irrespective of the electron sources and spatial transport
mode. To delineate most clearly the consequences of the Klein-Nishina
drops in the radiative loss rate, we chose as illustrative example the
simplest realistic model for \kr electron dynamics in the Galaxy,
consisting of the competion of radiative losses and secondary
production by inelastic hadron-hadron collisions. We demonstrate that
the spectral structure in the FERMI and H.E.S.S. data is well
described and even the excess measured by ATIC might be explained by
Klein-Nishina steps.

After completing this work we noticed the recent preprint by Stawarz,
Petrosian and Blandford (2009) who also explain the recently measured
galactic electron spectrum by the Klein-Nishina suppression of the
inverse Compton energy loss of relativistic electrons in an optical
photon field with an energy density of 3 eV cm$^{-3}$.
\begin{acknowledgements} 
This work was partially supported by the German Ministry for Education
and Research (BMBF) through Verbundforschung Astroteilchenphysik grant
05 A08PC1 and the Deutsche Forschungsgemeinschaft through grant Schl
201/20-1.
\end{acknowledgements}

\begin{appendix}
\section{Inverse Compton energy loss rate in graybody photon distributions}
The inverse Compton power of a single electron in a general target
photon field $n(\ep )$ is (Ch. 4.2 in \cite{s02})
\be
p_C(\es , \gamma )=c\int_0^\infty d\ep n(\ep )\es \sigma (\es ,\ep , \gamma )~,
\label{b1}
\ee
where $\es $ denotes the scattered photon energy. The differential
Klein-Nishina cross section\cite{bk70} is given by
\be
\sigma (\es ,\ep , \gamma )={3\sigma _T\over 4\ep \gamma ^2}G(q, \Gamma )
\label{b2}
\ee
with 
\begin{align}
G(q, \Gamma )=&\,G_0(q)+{\Gamma ^2q^2(1-q)\over 2(1+\Gamma q)}
\nonumber\\
\textnormal{where~~}&\,G_0(q)=2q\ln q +(1+2q)(1-q)
\label{b3}
\end{align}
and
\be
\Gamma ={4\ep \gamma \over mc^2},\;\;~~~ q={\es \over \Gamma (\gamma mc^2-\es )}~.
\label{b4}
\ee
By integrating over all kinematically allowed scattered photon
energies we find for the inverse Compton energy loss rate of a single
electron
\begin{align}
|\dot{\gamma }|_C&={1\over mc^2}\int_0^{\epsilon _{s, \rm max}}d\es p_{\rm C}(\es , \gamma )
\nonumber\\
&={3c\sigma _T\over 4mc^2\gamma ^2}\int_0^\infty d\ep \ep ^{-1}n(\ep )
\int_0^{\epsilon _{s, \rm max}}d\es \es G(q, \Gamma )~,
\label{b5}
\end{align}
where $\epsilon _{s, \rm max}=\Gamma \gamma mc^2/(\Gamma +1)$
corresponds to $q=1$. Using $q$ as integration variable instead of
$\es $ results in
\begin{align}
|\dot{\gamma }|_{\rm C}&={12c\sigma _T\over m_ec^2}\gamma ^2\int_0^\infty d\ep \ep n(\ep )J(\Gamma )
\nonumber\\
&\textnormal{with~~}J(\Gamma )=\int_0^1dq\, {qG(q,\Gamma )\over (1+\Gamma q)^3}~.
\label{b6}
\end{align}
For the graybody photon distribution 
\be
n_G(\ep )={15W\over \pi ^4(k_BT)^4}{\ep ^2\over \exp \l[\ep /k_BT\r]-1}
\label{b7}
\ee
the inverse Compton energy loss rate then is 
\begin{align}
|\dot{\gamma }|_{\rm C}&={20c\sigma _TW\over \pi ^4m_ec^2}\gamma ^2I(\gamma ,T)
\nonumber\\
\textnormal{with~~}&I(\gamma ,T)=9(k_BT)^{-4}\int_0^\infty d\ep\, {\ep ^3\over  \exp \l[\ep /k_BT\r]-1}J(\Gamma ) ~.
\label{b8}
\end{align}
Jones\cite{j65} already noted that the double integral $I(\gamma ,T)$
cannot be solved exactly, so that approximations (see
e.g. \cite{pe09}) are required. It has been noted\cite{s09} that the
integral $J(\Gamma )$ is reasonably well approximated by
\be
J(\Gamma )=\int_0^1dq\, {qG(q,\Gamma )\over (1+\Gamma q)^3}\simeq {1\over 9+2\Gamma ^2}, 
\label{b9}
\ee
so that the double integral in Eq. (\ref{b8}) becomes 
\be
I(\gamma ,T)=(k_BT)^{-4}\int_0^\infty d\ep \, {\ep ^3\over  \exp \l[\ep /k_BT\r]-1}
{1\over 1+{32\gamma ^2\ep ^2\over 9m_e^2c^4}}~.
\label{b10}
\ee
With the substitution $x=\ep /k_BT$ we find 
\be
I(A)=\int_0^\infty dx \, {x^3\over 1+{x^2\over A^2}}{1\over e^x-1}~,
\label{b11}
\ee
where 
\be
A={3m_ec^2\over \sqrt{32}}k_BT\gamma~.
\label{b12}
\ee
The series 
\bdm 
{1\over 1+{x^2\over A^2}}=\sum_{k=1}^\infty (-1)^{k-1}{x^{2(k-1)}\over A^{2(k-1)}}
\edm
leads to 
\be
I(A)=\sum_{k=1}^\infty (-1)^{k-1}{\Gamma \l[2k+2\r]\zeta \l[2k+2\r]\over A^{2(k-1)}},
\label{b13}
\ee
which, for $A\ge 1$ to lowest order in $A^{-2}$, yields 
\be
I(A\ge 1)\simeq \Gamma \l[4\r]\zeta \l[4\r]={\pi ^4\over 15}~.
\label{b14}
\ee
For $A<1$ we approximate the integral (\ref{b11}) by 
%
\begin{align}
I(A<1)&\simeq \int_0^Adx\, {x^3\over e^x-1}+A^2\int_A^\infty dx\, {x\over e^x-1}
\nonumber\\
&\simeq A^2\int_0^\infty dx\, {x\over e^x-1}\, +\int_0^Adx\, x^2\, -A^3
\nonumber\\
&={\pi ^2A^2\over 6}-{2A^3\over 3}\simeq {\pi ^2A^2\over 6}
\label{b15}
\end{align}
We combine the two expansions (\ref{b14}) and (\ref{b15}) to 
\be
I(A)\simeq {\pi ^4\over 15}{1\over 1+{2\pi ^2\over 5A^2}}
\label{b16}
\ee
valid at all values of $A$. The approximation (\ref{b16}) can be written as 
\be
I(A)\simeq {\pi ^4\over 15}{1\over 1+\l({\gamma \over \gamma _K}\r)^2}~,
\label{b17}
\ee
where we introduce the critical Klein-Nishina Lorentz factor 
\be
\gamma _{K}\equiv {3\sqrt{5}\over 8\pi }{m_ec^2\over k_BT}={0.27m_ec^2\over k_BT}~.
\label{b18}
\ee
Using this approximation in Eq. (\ref{b8}) readily yields the inverse
Compton loss rate (\ref{a2}).

\end{appendix}

\end{document}